\pdfoutput=1
\documentclass[sigconf,natbib=true,anonymous=false]{acmart}

\usepackage[ruled,vlined]{algorithm2e}
\usepackage{subfigure}
\usepackage{multirow}
\usepackage[normalem]{ulem}

\usepackage{amsfonts,amssymb}
\usepackage{amsmath,bm}
\usepackage{threeparttable}
\usepackage{enumitem}
\usepackage{color}
\usepackage{hyperref}
\usepackage{bbm}
\hypersetup{colorlinks=false,linkcolor=blue,urlcolor=blue,citecolor=red}
\usepackage{balance}

\newcommand{\Mat}[1]{\textbf{#1}}

\newcommand{\Space}[1]{\mathbb{#1}}
\newcommand{\Set}[1]{\mathcal{#1}}

\newcommand{\ie}{\emph{i.e., }}
\newcommand{\eg}{\emph{e.g., }}

\newcommand{\wrt}{\emph{w.r.t. }}

\newcommand{\ourmethod}{KRDN\xspace}

\AtBeginDocument{%
  \providecommand\BibTeX{{%
    \normalfont B\kern-0.5em{\scshape i\kern-0.25em b}\kern-0.8em\TeX}}}

\copyrightyear{2023}
\acmYear{(\textit{Corresponding author: Yuren Mao}). 2023}
\setcopyright{acmlicensed}\acmConference[SIGIR '23]{Proceedings of the 46th International ACM SIGIR Conference on Research and Development in Information Retrieval}{July 23--27, 2023}{Taipei, Taiwan}
\acmBooktitle{Proceedings of the 46th International ACM SIGIR Conference on Research and Development in Information Retrieval (SIGIR '23), July 23--27, 2023, Taipei, Taiwan}
\acmPrice{15.00}
\acmDOI{10.1145/3539618.3591707}
\acmISBN{978-1-4503-9408-6/23/07}

\begin{document}
\fancyhead{}

\title{Knowledge-refined Denoising Network for \\ Robust Recommendation}

%\author{Xinjun Zhu$^{\mathsection}$, Yuntao Du$^{\dagger}$, Yuren Mao$^{\mathsection}$, Lu Chen$^{\dagger}$, Yujia Hu$^{\dagger}$, and Yunjun Gao$^{\dagger}$}
%\affiliation{
 %{\large$^{\dagger}$}College of Computer Science, Zhejiang University, Hangzhou, China\\
 %{\large$^{\mathsection}$}School of Software, Zhejiang University, Ningbo, China\\
 %\{$^{1}$xjzhu, $^{2}$ytdu, $^{3}$yuren.mao, $^{4}$luchen, $^{5}$charliehu, %$^{6}$gaoyj\}@zju.edu.cn
 %\country{}}

\author{Xinjun Zhu}
\affiliation{%
    \institution{Zhejiang University}
    \country{}
}
\email{xjzhu@zju.edu.cn}

\author{Yuntao Du}
\affiliation{%
    \institution{Zhejiang University}
    \country{}
}
\email{ytdu@zju.edu.cn}

\author{Yuren Mao}
\affiliation{%
    \institution{Zhejiang University}
    \country{}
}
\email{yuren.mao@zju.edu.cn}

\author{Lu Chen}
\affiliation{%
    \institution{Zhejiang University}
    \country{}
}
\email{luchen@zju.edu.cn}

\author{Yujia Hu}
\affiliation{%
    \institution{Zhejiang University}
    \country{}
}
\email{charliehu@zju.edu.cn}

\author{Yunjun Gao}
\affiliation{%
    \institution{Zhejiang University}
    \institution{Alibaba-Zhejiang University Joint Institute of Frontier Technologies}
    \country{}
}
\email{gaoyj@zju.edu.cn}

\renewcommand{\shortauthors}{Zhu, et al.}

\begin{abstract}

Knowledge graph (KG), which contains rich side information, becomes an essential part to boost the recommendation performance and improve its explainability. However, existing knowledge-aware recommendation methods directly perform information propagation on KG and user-item bipartite graph, ignoring the impacts of \textit{task-irrelevant knowledge propagation} and \textit{vulnerability to interaction noise}, which limits their performance. To solve these issues, we propose a robust knowledge-aware recommendation framework, called \textit{Knowledge-refined Denoising Network} (KRDN), to prune the task-irrelevant knowledge associations and noisy implicit feedback simultaneously. KRDN consists of an adaptive knowledge refining strategy and a contrastive denoising mechanism, which are able to automatically distill high-quality KG triplets for aggregation and prune noisy implicit feedback respectively. Besides, we also design the self-adapted loss function and the gradient estimator for model optimization. The experimental results on three benchmark datasets demonstrate the effectiveness and robustness of KRDN over the state-of-the-art knowledge-aware methods like KGIN, MCCLK, and KGCL, and also outperform robust recommendation models like SGL and SimGCL. The implementations are available at \url{https://github.com/xj-zhu98/KRDN}.

\end{abstract}




\maketitle

\begin{figure}[t]
	\centering
        \vspace{5pt}
	\includegraphics[width=0.43\textwidth]{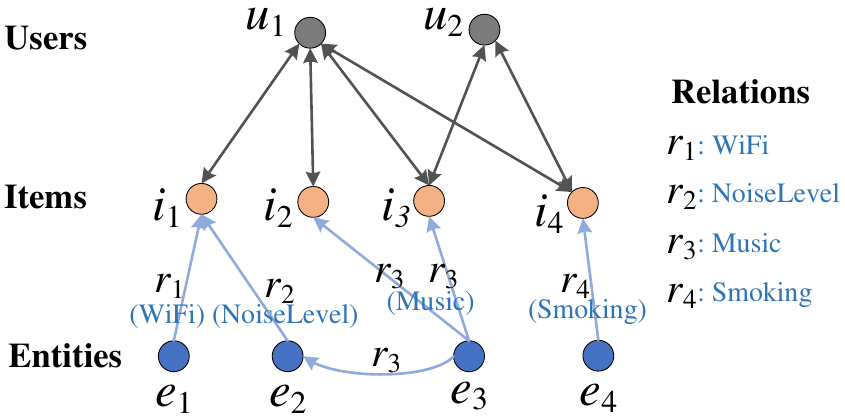}
	\vspace{-10pt}
	\caption{An example of knowledge-aware recommendation on Yelp2018 dataset.}
	\label{fig:example}
 	\vspace{-15pt}
\end{figure}

\section{Introduction}
\label{sec:introduction}

In the Internet era of information explosion, recommendation systems are widely deployed in real-life applications such as E-commerce, online advertisement, and social media platforms to provide personalized information services. Traditional recommendations (\eg collaborative filtering~\cite{kdd08cf,icdm08cf,sigir19ngcf,www23ecf}) heavily rely on historical user interaction data, which brings data sparsity and cold-start problems; furthermore, their vulnerability to interaction noise also degrades the recommendation performance~\cite{wsdm2021T-CE, www2022DeCA, mm2021IR}. Recently, knowledge graph (KG), which provides rich side information among items, has demonstrated great potential in alleviating cold-start issues and improving the robustness and explainability of recommendations.

Incorporating external knowledge from KG to learn high-quality user and item representations has become the concept of knowledge-aware recommendation. Early work~\cite{kdd16cke,www18dkn,alg18cfkg} on this topic directly integrates knowledge graph embeddings with items to enhance representations. Then, to further improve the performance of knowledge-aware recommendation, knowledge meta-paths-based methods are proposed~\cite{kdd18metapath,aaai19explainable,kdd20hinrec}, which enriches the interactions with meta-paths from users to items for better exploiting user-item connectivities. However, due to the difficulty of obtaining informative meta-paths, these methods suffer from labor-intensive process~\cite{aaai19explainable}, poor scalability~\cite{kdd18metapath}, and unstable performance~\cite{sigir19reinforceKG}. To address these issues, Graph Neural Networks (GNNs)~\cite{iclr17gcn,nips17sageGCN,iclr18gat} are adopted in knowledge-aware recommendation to achieve end-to-end recommendation by means of iteratively propagating high-order information over KG\cite{cikm18ripple,kdd19kgat,kdd19kgnn-ls,sigir20ckan,cikm21kcan,www21kgin,sigir22kgcl,sigir22mcclk}. These propagation-based methods can effectively gather multi-hop neighbors into representations, which enable to achieve the impressive performance of recommendation.

% argue drawbacks, different contribution of facts, ignore using facts to enhance robustness
% two figures, one for fine-grained facts, another for noisy message passing
Despite effectiveness, we argue current propagation-based methods commonly have the following two limits:
\vspace{3pt}
\begin{itemize}[leftmargin=*]
    % or Coarse-grained facts modeling
    % \item \textbf{Coarse-grained KG modeling}. To the best of our knowledge, none of these works consider modeling KG triplets at a finer-grained level of facts. An important \todo{evidence} has been long ignored: due to the generalization and scale of the knowledge graph, facts in KG may contribute differently to recommendation, according to their relatedness with items. \todo{Taking the left of \todo{Figure X} as an example, fact x-> y and y <- z indicates that book x and book z are written by the same author y, which reveal important semantics beyond collaborative signals. However, other facts like author is a friend of author b (a <-> b ) can be negligible, because these information is far related to our recommendation targets.}
    \item \textbf{Task-irrelevant Knowledge Propagation.} Previous research blindly aggregate all kinds of information in KG into item representations, regardless of their semantic relatedness with recommendation task. However, due to the scale and generalization of knowledge graphs~\cite{aaai15transR}, they can be rather noisy, and some facts in KG are semantically far away from recommendation scenarios. Taking the Figure~\ref{fig:example} as an example, the three businesses $i_1, i_2, i_3$ interacted by $u_1$, where $i_2, i_3$ are resident in \textit{music} $(r_{3})$, $i_1$ is associated with the \textit{NoiseLevel} $(r_{2})$. It can be concluded that $u_1$ focuses on the ambience, while \textit{WiFi} $(r_{1})$ is a more marginal attribute linked to $i_1$. Integrating these irrelevant facts such as $(i_1, r_1, e_1)$ are not useful for learning high-quality user and item representations, and can also introduce unnecessary noise and thus degrade recommendation performance.
    % Need to illustrate why previous studies fail to consider this? (answered in related work)
    % keep short and straightforward to our work
    % sequential recommendation with GNN / mask / denoising 论文参考
    \item \textbf{Vulnerable to Interaction Noise}. To align with the graph structure of KG, existing studies typically construct user-item interaction graph from implicit feedback and propagate collaborative information with GNNs. However, the recursive message passing scheme of GNNs is known to be vulnerable to the quality of the input graphs~\cite{sigir21mask}, and implicit feedback is inherently noisy~\cite{wsdm2021T-CE, sigir22SGDL}. Directly performing propagation on such a noisy interaction graph would make the model difficult to learn users' real interests and degrades the performance. For example, $u_1$ and $i_4$ are related in structure, maybe because $u_1$ and $u_2$ have both been to $i_3$, the recommendation system recommends $i_4$ to $u_1$ according to the transient behavior similarity between the two users, and $u_1$ happens to have an interaction with $i_4$. However, from the knowledge information brought by KG, $i_1,i_2,i_3$ that $u_1$ interacts with are all environment-conscious businesses, while $i_4$ supports \textit{smoking} $(r_4)$, which is inconsistent with $u_1$'s preference. So, $u_1$ is semantically distinct from $i_4$, which is probably a noisy interaction. Therefore, ignoring the noisy interactions would propagate misleading messages and contaminate the entire graph.
    % Previous methods typically use fixed user-item graph to propa-gate collaborative information
     %~\cite{mm20grcn,mm21denoise}.
    % Should introduce another noise: false negative behaviors?
\end{itemize}

\vspace{3pt}
% To tackle the aforementioned challenges, we carefully explore the semantic relatedness in KG for fine-grained knowledge propagation, and also propose a novel denoising scheme to prune noisy interactions with the help of KG. Specifically, we propose a new model, \underline{K}nowledge-\underline{R}efined \underline{D}enoising \underline{N}etwork (\ourmethod), which can not only make full use of relevant knowledge in KG to promote recommendation performance, but also show good robustness to noisy implicit interactions. \ourmethod consists of two components to correspondingly address the above problems:
To tackle the aforementioned challenges, we carefully explore the conduction of knowledge and collaborative signals on KG and user-item graphs, and then propose a novel denoising scheme to prune the task-irrelevant knowledge and noisy interactions simultaneously. Specifically, we propose a new model, \underline{K}nowledge-\underline{R}efined \underline{D}enoising \underline{N}etwork (\ourmethod), which can not only make full use of relevant knowledge in KG to promote recommendation performance, but also show excellent robustness to noisy implicit interactions. \ourmethod consists of two components to correspondingly address the above problems:

% briefly introduce the core idea
\vspace{3pt}
\begin{itemize}[leftmargin=*]
    \item \textbf{Adaptive Knowledge Refining.} An adaptive pruning strategy is proposed to distill high-quality triplets and offer additional knowledge for the recommendation, which can be jointly optimized with downstream recommendation tasks. Besides, according to their relatedness with items, each KG triplet is recognized with a certain type of facts (\ie ``item-item'' facts, ``item-attribute'' facts, and ``attribute-attribute'' facts, detailed in Section~\ref{sec:compositional-knowledge-aggregation}). Based on pruned knowledge and multi-faceted facts, we design a novel compositional knowledge aggregation mechanism to effectively capture refined and multi-faceted contexts into representations for better characterizing items.
    
    % \item \textbf{Contrastive Denoising Learning.} To avoid noisy message passing and improve the robustness of recommendation, we investigate the semantic similarity of items from both collaborative signals and knowledge associations, and devise a constrastive denoising mechanism to identify noisy interactions for learning user true preference.  \todo{Specifically, \ourmethod iteratively adjust weights of possible noisy edges in terms of the contrastive item similarity knowledgable semantics of items. }
    \item \textbf{Contrastive Denoising Learning.} To avoid noisy message passing and improve the robustness of recommendation, we investigate the collaborative and knowledge similarities and devise a contrastive denoising mechanism to capture the divergence between them and identify noisy interactions for learning user true preference. Specifically, \ourmethod iteratively adjusts weights of possible noisy edges through a relation-aware self-enhancement mechanism in both soft and hard manners.
\end{itemize}

% effectiveness of our model 
% To this end, the newly proposed \ourmethod model is designed to i) exploit the auxiliary knowledge information in a fine-grained manner, and ii) better capture user true preference by denoising interactions with the help of KG. We conduct extensive experiments on three real-world datasets to evaluate the performance of \ourmethod and existing methods. Experimental results show that our \ourmethod significantly outperforms all the start-of-the-art methods such KGIN~\cite{www21kgin}, KGCL~\cite{sigir22kgcl}, MCCLK~\cite{sigir22mcclk} and SimGCL~\cite{sigir22simgcl}. Furthermore, \ourmethod is able to identify noise interactions and much robust than other methods. 
To this end, the newly proposed \ourmethod model is designed to i) adaptively refine knowledge associations, and ii) better capture users' true preferences by denoising interactions with the help of KG. We conduct extensive experiments on three real-world datasets to evaluate the performance of \ourmethod and existing methods. Experimental results show that our \ourmethod significantly outperforms all the start-of-the-art methods such KGIN~\cite{www21kgin}, KGCL~\cite{sigir22kgcl}, MCCLK~\cite{sigir22mcclk} and SimGCL~\cite{sigir22simgcl}. Furthermore, \ourmethod is able to identify noisy edges and more robust than other methods.

In summary, our contributions are as follows:

\begin{itemize}[leftmargin=*]
    % \item We approach knowledge-aware recommendation from a new perspective by taking the fine-grained knowledge and denoising collaborative signals into consideration.
    \item We approach knowledge-aware recommendation from a new perspective by refining knowledge associations and denoising implicit interactions simultaneously.
    % \item We exploit the knowledge association in KG in a fine-grained way, which can not only learn to prune irrelevant triplets in the light of collaborative signals, but also aggregate different kinds of facts compositionally for high-quality knowledge representations.
    \item This paper exploits the knowledge association in KG in a fine-grained way, which can not only learn to prune irrelevant triplets in the light of downstream supervision signals, but also aggregate multi-faceted facts compositionally for high-quality knowledge representations.
    % \item We propose a contrastive denoising strategy by leveraging the semantic similarities from both collaborative and knowledge aspects to better represent and propagate user true preference, which can greatly enhance the robustness of recommendations.  
    \item We propose a contrastive denoising strategy by leveraging the semantic divergence between collaborative and knowledge aspects to better represent and propagate user true preference, which can greatly enhance the robustness of recommendations.  
    \item Extensive experiments on three public benchmark datasets are conducted to demonstrate the superiority of \ourmethod.
\end{itemize}

% we use XX and XXX interchangeably in this work.

% 介绍现在 KG 的缺点

% less informative entities and relations

% noise

% 介绍现在 denoise 的方法
% not all the information from high-order neighbors are positive in reality~\cite{www21imp-gcn}.

% KG 被用来作为辅助数据，提供semantic信息
% 但 KG 本身存在的问题：KG过大，很多neighbor对推荐的item没有效果，attention的weight decay问题
% 同时，KG 的语义信息可用来对 CF 中的 noise 进行 denoise

% \subsection{Misc}

% Intrinsic graph-structures are not always optimal for the downstream tasks~\cite{nips2020iterative}

% NeuralSparse~\cite{icml20ns} learns k-neighbor subgraphs for robust
% graph representation learning by selecting at most k edges for
% each nodes. The k-neighbor assumption however limits its learning
% power and may lead to suboptimal performance in generalization

% Using Gumbel-Softmax based approaches
% in a sequential setting is difficult as the bias accumulates because of mixing errors~\cite{aaai2018softmax}

% Most GNN methods are highly sensitive to the quality of graph structures and usually require a perfect graph structure for learning informative embeddings. Since GNNs compute node embeddings by recursively

% % negative sampling 
% most negatives are easily separable, which are not necessary.

% % 这里要说和 www20 xiang wang 论文的区别
\section{Problem Formulation}
\label{sec:problem_formulation}

We begin by introducing structured data relating to our investigated problem, and then we formulate our task.

\vspace{3pt}
\noindent
\textbf{User-Item Bipartite Graph.}
In this paper, we concentrate on inferring the user preferences from the implicit feedback~\cite{uai09BPRloss}. To be specific, the behavior data (\eg click, comment, purchase) involves a set of users $\mathcal{U} = \{u\}$ and items $\mathcal{I} = \{i\}$. We view user-item interactions as a bipartite graph $\mathcal{G}_b$, and construct the interaction matrix $\mathbf{R} \in \mathbb{R}^{|\mathcal{U}| \times|I|}$, where $|\mathcal{U}|$ and $|\mathcal{I}|$ denote the number of users and items, respectively. Each entry $\mathbf{R}_{ui} = 1$ if user $u$ has interacted with item $i$, and $\mathbf{R}_{ui} = 0$ otherwise. Note that implicit feedback is inherently noisy~\cite{icdm08cf,uai09BPRloss}, and observed interactions are not necessarily positive~\cite{sigir21enhanced, mm20grcn}, which would lead to sub-optimal performance.  We will discuss how to address the problem by leveraging complementary information of KG in Section~\ref{sec:denoising_cf}. 

% Its corresponding adjacency matrix $\mathbf{A}$ for the bipartite
% graph can be defined as $A=\left[\begin{array}{cc}0 & \mathrm{R} \\ \mathrm{R}^{\top} & 0\end{array}\right]$, which 

\vspace{3pt}
\noindent\textbf{Knowledge Graph.}
KGs hold structured data about real-world facts, like item attributes, concepts, or external commonsense. Let KG be a heterogeneous graph $\mathcal{G}_k= \{(h,r,t) | h,t\in \mathcal{E}, r\in \mathcal{R}\}$,  where each triplet $(h,r,t) \in \mathcal{T}$ means that a relation $r$ exists from head entity $h$ to tail entity $t$; $\mathcal{T}$, $\mathcal{E}$ and $\mathcal{R}$ refer to the sets of triplets, entities, and relations in $\mathcal{G}_k$. For example, a triple (\textit{Mark Hamill}, \textit{ActorOf}, \textit{Star War}) indicates that \textit{Mark Hamill} is an \textit{actor} of the movie \textit{Star War}. Therefore, we can link items with entities ($\mathcal{I} \subset \mathcal{E}$) to offer auxiliary semantics to interaction data. However, observations in Figure~\ref{fig:example} show that KG involves numerous task-irrelevant triplets, which cause a serious impact on recommendation. We will demonstrate how to model fine-grained facts over KG in Section~\ref{sec:facts_modeling}.

% 将 fine-grained facts 放到相应的地方去讲比较好?
% we denote triplets as "item-item" facts if both entities are aligned with items; "item-attr" (short for "item-attribute") facts represent that one of the entities is associate with item while the other acts as the attribute of it; "attr-attr" (short for "attribute-attribute") facts stand for those triplets that are not directly related to items.
% \vspace{3pt}
% \noindent\textbf{Fine-grained Facts.} We categorize KG triplets $\mathcal{T}$ into three disjoint subsets $\mathcal{T} = \{\mathcal{T}_{1}, \mathcal{T}_{2},\mathcal{T}_{3} \}$, where $\mathcal{T}_{1} = \{(h,r,t) | h,t \in \mathcal{I}\}$ denotes as "item-item" facts that both entities in triplet are aligned with items;  $\mathcal{T}_{2} = \{(h,r,t) | h \in \mathcal{I} \text{or } t \in \mathcal{I} \}$ denotes "item-attr" facts, which means that one of the entities is related to item while the other acts as the attribute of it; $ \mathcal{T}_{3} = \{(h,r,t) | h,t \notin \mathcal{I} \}$ denotes "attr-attr" facts that both entities are represented as attributes. As a result, we reorganize KG triplets with multiple facts, which differs from the unified knowledge modeling adopted in previous works~\cite{kdd19kgat,sigir20ckan,www21kgin}.

\vspace{3pt}
\noindent\textbf{Task Description.}
Given the user-item bipartite graph $\mathcal{G}_b$ and the KG $\mathcal{G}_k$, our task of knowledge-aware recommendation is to predict how likely that a user would adopt an item which he has not engaged with before.

\section{Methodology}
\label{sec:methodology}

\begin{figure*}[t]
    \vspace{-20pt}
	\centering
	\vspace{-15pt}
	\includegraphics[width=0.90\textwidth]{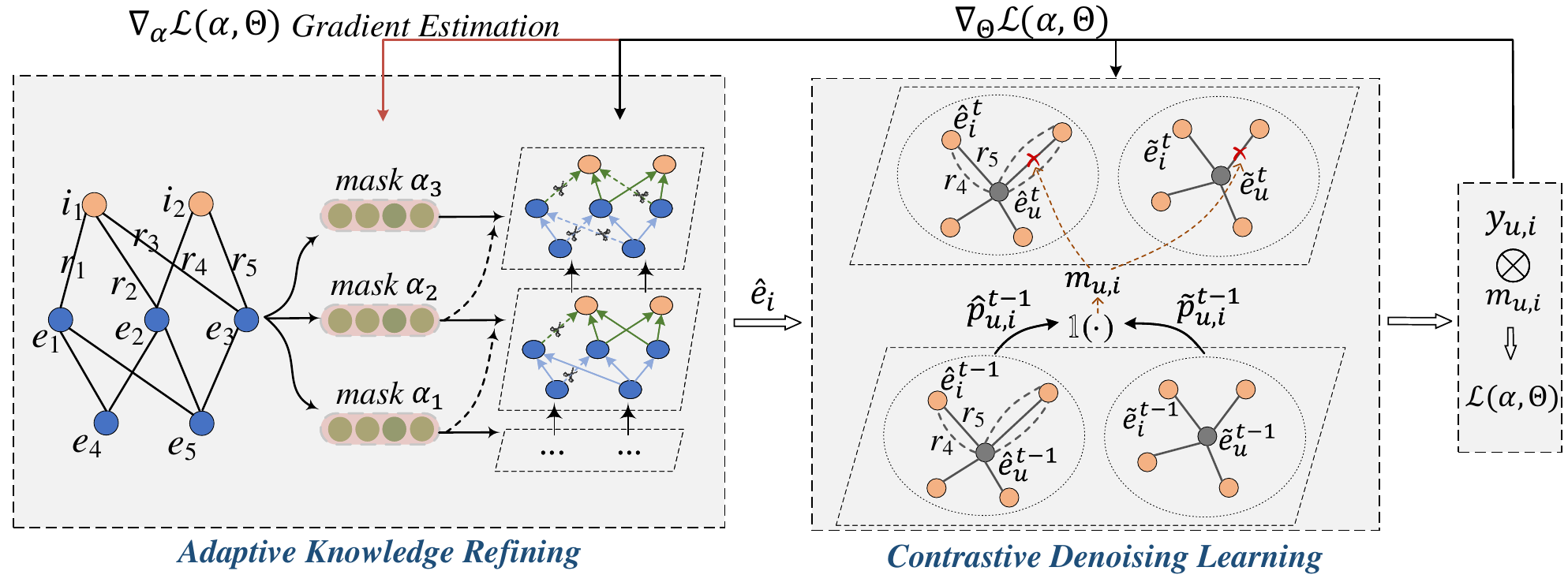}
	\vspace{-10pt}
	\caption{Illustration of the proposed \ourmethod framework. Interrupt task-irrelevant knowledge propagation in KG with multiple masks (left). Identify noise interactions according to the divergence between collaborative and knowledge signals (right).}
	\label{fig:overview}
	\vspace{-10pt}
\end{figure*}

We present the proposed architecture of \ourmethod. Figure~\ref{fig:overview} shows the model framework, which consists of two key components: (1) adaptive knowledge refining, which uses parameterized binary masks to learn to remove irrelevant facts with an unbiased gradient estimator, meanwhile designs a compositional knowledge aggregator to effectively integrate different kinds of knowledge associations for contextual propagation; and (2) contrastive denoising learning, which focuses on the difference of collaborative and knowledge signals to identify noisy interactions in a contrastive way and iteratively performs relation-aware graph self-enhancement to augment user representations.

\subsection{Adaptive Knowledge Refining}
\label{sec:facts_modeling}
% introduce fine-grained facts
Unlike previous propagation-based methods~\cite{www21kgin,sigir22mcclk,sigir22kgcl} that directly integrate all kinds of information in KG into item representations, we aim to capture the most relevant knowledge associations which are beneficial for learning user preference. Specifically, we design an adaptive pruning mechanism that learns to prune extraneous facts with trainable stochastic binary masks, and devise a gradient estimator to jointly optimize them with the model. 

Meanwhile, we argue that existing approaches are unable to characterize items properly because they do not differentiate item-related knowledge associations from the rest, and only aggregate KG information at a \textit{coarse} granularity. Different entities in KG have different semantics for recommendation scenarios, and they play different roles in profiling items. This motivates us to perform a context-aware compositional aggregation mechanism to gather different semantics according to the relatedness of items.

\subsubsection{\textbf{Irrelevant Facts Pruning.}} 
% 先要说明为什么我们要这么做的动机，同时指出现有相似做法的不妥之处
% 提一下对比KGCL的优势，1、不需要构造多个view；2、每一层GNN聚合都可以进行pruning，multi-hop information pruning noisy association
As we discussed in Section~\ref{sec:introduction}, KG contains lots of noisy and task-irrelevant information, which is not useful or even degrades the performance. One straightforward solution is to manually construct a K-neighbor subgraph to constrain the receptive fields of nodes~\cite{sigir20atbrg,cikm21kcan}, or randomly drop some edges to construct multi-view graph structure for contrastive learning~\cite{sigir22kgcl,sigir22mcclk}. However, these approaches highly rely on the quality of graph construction, and cannot adaptively drop unnecessary edges according to the recommendation tasks. Hence, we turn to a parameterized method to jointly learn the optimal pruning strategy with downstream collaborative signals. Technically, we first attach each triplet in $\mathcal{T}$ with a binary mask $m \in \{0,1\}$ to indicate whether the triplet should be dropped, so the post-pruned facts can be expressed as:
\begin{equation}
\tilde{\mathcal{T}}= \{(h_i, r_i, t_i) | m_i = 1\}
\end{equation}
where $\tilde{\mathcal{T}}$ is the subset of $\mathcal{T}$. However, directly optimizing the masks $M$ is computationally intractable due to its discrete, non-differentiability and combinatorial nature of $2^{|\mathcal{T}|}$ possible states~\cite{iclr2016gumbel-softmax,nips20disARM}. To address this challenge, we consider each $m_i$ is subject to a Bernoulli distribution with parameter $\sigma(\alpha_i)$, so that
\begin{equation}
m_i \sim \operatorname{Bern}(\sigma(\alpha_i))
\end{equation}
where we choose the widely used sigmoid function as the deterministic function $\sigma(\cdot)$, so that the parameters $\alpha$ can be bounded with $(0,1)$. To let masks $M$ be jointly optimized with recommendation task, we integrate them with our target loss $\mathcal{L}$, and reformulate with Bernoulli parameters as:
\begin{equation}
\label{eq:elbo}
\mathcal{\tilde{L}}(\alpha,\Theta) = \mathbb{E}_{M \sim \prod_{i=1}^{K} \operatorname{Bern} (m_i; \alpha_i)} [\mathcal{L}(M,\Theta)]
\end{equation}
where $\mathbb{E}$ is the expectation, $\Theta$ denotes the rest parameters of models, and $\mathcal{\tilde{L}}$ is the evidence lower bound\footnote{This can be derived by the Jensen's Inequality.} (ELBO)~\cite{1999elbo} for objective $\mathcal{L}$ over the parameters $\alpha$. To minimize the expected cost via gradient descent, we need to estimate the
gradient $\nabla_{\alpha} \mathcal{\tilde{L}}(\alpha,\Theta)$. Note that there are several studies have been proposed to estimate the gradients for discrete variables, such as REINFORCE~\cite{1992reinforce}, Gumbel-Softmax~\cite{iclr2016gumbel-softmax}, straight-through~\cite{arxiv13Straight}, hard concrete~\cite{iclr2017l_0} and ARM~\cite{iclr2018arm}. However, those approaches either suffer from high variance or biased gradients. Thus, we adopt DisARM~\cite{nips20disARM}, a recently proposed unbiased and low-variance gradient estimator, to efficiently backpropagate the gradient of parameters $\alpha$. We will introduce masks optimization in Section~\ref{sec:disarm}

% 介绍如何 prune 不需要的 facts
% 介绍 composition-based aggregation
\subsubsection{\textbf{Compositional Knowledge Aggregation.}}\label{sec:compositional-knowledge-aggregation}
To better understand the semantic relatedness of KG triplets, we categorize $\mathcal{T}$ into three disjoint subsets $\mathcal{T} = \{\mathcal{T}_{1}, \mathcal{T}_{2},\mathcal{T}_{3} \}$, in terms of their connectivities to items. Specifically, we denote $\mathcal{T}_{1} = \{(h,r,t) | h,t \in \mathcal{I}\}$ as ``item-item'' facts, as both entities in these triplets are aligned with items. Similarly, $\mathcal{T}_{2} = \{(h,r,t) | h \in \mathcal{I} \text{or } t \in \mathcal{I} \}$ stands for ``item-attribute'' facts, which means that one of the entities is related to item while the other acts as the attribute of it. The rest triplets $ \mathcal{T}_{3} = \{(h,r,t) | h,t \notin \mathcal{I} \}$ are ``attribute-attribute'' facts, where both entities are represented as attributes. As a result, we reorganize KG triplets with multi-facet facts, which can explicitly gather different knowledge associations. To denoise on message passing, we propose a new aggregation mechanism consisting of noisy message pruning and compositional knowledge aggregation. Specifically, we use $\mathcal{N}_{h}=\{(r,t) | (h,r,t)\in \Set{G}_k\}$ to represent the neighborhood entities and the first-order relations of item $h$ in KG, and propose to integrate the multi-faceted relational context from neighborhood entities to generate the \textit{knowledge representation} of entity $h$:
{\setlength{\abovedisplayskip}{3pt}
\setlength{\belowdisplayskip}{3pt}
\begin{equation}
\mathbf{e}_h^{(1)}=\frac{1}{|\mathcal{N}_h|} \sum_{(r, t) \in \mathcal{N}_h} \operatorname{ReLU}\big(\mathbf{W} \mathcal{\phi}(\mathbf{e}_t^{(0)}, \mathbf{e}_r)\big) \cdot m_{h, t}^{(0)}
\end{equation}}where $\operatorname{ReLU}$ is the activation function, $m_{h, t}^{(0)} \in \{0, 1\}$ denotes whether triplet $(h,r,t)$ should be pruned or not, and $\phi(\cdot)$ is the aggregation function which gathers information from neighboring entities and corresponding relations. To differentiate different facts when aggregation, we design a compositional knowledge aggregator to integrate three kinds of semantics for avoiding interference between disparate information channels as follows:
\begin{equation}
\mathbf{W} \phi\left(\mathbf{e}_t, \mathbf{e}_r\right)=\left\{\begin{array}{l}
\mathbf{W}_{1}\left(\mathbf{e}_t \odot \mathbf{e}_r\right),(h, r, t) \in \mathcal{T}_{1,3} \\
\mathbf{W}_{2}\left(\mathbf{e}_t \oplus \mathbf{e}_r\right),(h, r, t) \in \mathcal{T}_2
\end{array}\right.
\end{equation}
where we use a relational-aware aggregation scheme~\cite{www21kgin} for same-level facts (``item-item'' facts and ``attribute-attribute'' facts), and utilize additional operations for cross-level facts (``item-attribute'' facts). Besides, we also introduce two trainable transformation matrices $\mathbf{W}_{1}, \mathbf{W}_{2} \in \Space{R}^{d \times d}$ to align the hidden semantics before aggregating the heterogeneous information together.

We further stack more aggregation layers to explore the high-order knowledge associations for items. Technically, we recursively formulate the knowledge representations of item $h$ after $l$ layers as:
\begin{equation}
\mathbf{e}_h^{(l)}=\frac{1}{|\mathcal{N}_h|} \sum_{(r, t) \in \mathcal{N}_h} \operatorname{ReLU}\big(\mathbf{W} \mathcal{\phi}(\mathbf{e}_t^{(l-1)}, \mathbf{e}_r)\big) \cdot m_{h, t}^{(l-1)}
\end{equation}

\subsection{Contrastive Denoising Learning}
\label{sec:denoising_cf}
% 不需要进行structure上的增强，而是基于node embedding上的增强；我们不需要捕捉两个视图对应节点的相似性，而是致力于发掘对应节点经过语义增强后的不同。
% 1. 为什么要引入两个view; 2. 为什么用相对距离; 3. self-enhancement的作用

The second key component of \ourmethod framework is designed mainly to identify noisy interactions in user-item bipartite graph and propagate high-order discriminative collaborative signals to present user true preference. Existing methods~\cite{www21kgin,kdd19kgat,sigir20ckan} ignore the noise in interactions and directly aggregate all information from neighboring users/items, which would make the model difficult to differentiate between noise and user true preference and result in suboptimal user/item representations. More recently, some works~\cite{sigir22kgcl,sigir22mcclk} focus on constructing different graph views and utilize contrastive learning to enhance the robustness of recommendation models. Unfortunately, this approach would inevitably lose the structure information and fail to identify fake interactions explicitly. Thus, we aim to leverage the divergence between collaborative signals and knowledge associations to filter the noisy interaction in an end-to-end manner. We illustrate our approach in Figure~\ref{fig:overview}.

\subsubsection{\textbf{Denoising Collaborative Aggregation.}}
\label{sec:DCA}
% 1. 引入two-view; 2. 引入self-enhancement
The item representations generated from Section~\ref{sec:facts_modeling} contain refined knowledge associations, which are considered to have high confidence. By directly aggregating such item information from the interaction graph, we can obtain the knowledge representation of users and items. However, the user-item graph built upon implicit feedback inevitably contains noise~\cite{uai09BPRloss}. Therefore, we initialize additional item representations to capture the pure collaborative signal individually as a comparison. To prune the noisy interaction, we focus on the divergence between collaborative and knowledge signals. 

Since \ourmethod keeps the original graph structure instead of randomly perturbation~\cite{sigir22kgcl}, we are able to use the relative distance to assess the stability of $i$ with respect to the importance of $u$ as the basis for denoising:
\begin{equation}
\label{eq:binary_drop}
m_{u, i}=\mathbbm{1}\big(\lvert\sigma(\tilde{p}_{u, i})-\sigma(\hat{p}_{u, i})\rvert <\gamma \big)
\end{equation}
where $\mathbbm{1}(\cdot)$ is a binary indicator function that returns 1 if the condition is true otherwise returns 0, and $\gamma$ is a pre-defined threshold hyperparameter. $\tilde{p}_{u, i}$ and $\hat{p}_{u, i}$ respectively denote collaboration and knowledge similarities between $u$ and $i$, and $\tilde{p}_{u, i}$ can be simply formulated as:
\begin{equation}
\label{eq:p2}
\tilde{p}_{u, i}=\frac{\exp \left(s\left(\tilde{\mathbf{e}}_u, \tilde{\mathbf{e}}_i\right)\right)}{\sum_{i^{\prime} \in \mathcal{N}_{(u)}} \exp \left(s\left(\tilde{\mathbf{e}}_u, \tilde{\mathbf{e}}_{i^{\prime}}\right)\right)}
\end{equation}
where $\tilde{\mathbf{e}}_u$ and $\tilde{\mathbf{e}}_i$ are additional user and item representations, which are initialized to capture the pure collaborative signals, and $\mathcal{N}_{(u)}$ is used to represent the set of neighbors of node $u$ in the user-item graph. $s(\cdot)$ denotes the inner product to estimate the similarity. As for $\hat{p}_{u, i}$, item representations from KG involving multi-relation semantics, and user preference mixing various information via message passing in the graph. Hence, we estimate the correlation degree between user $u$ and item $i$ across multiple relations. Each item has a relation set noted $\mathcal{R}_{(i)} = \{r | (h,r,t) \in \mathcal{T} \text{and } h \in \mathcal{I} \}$, and the similarity can be formulated as follows:
\begin{equation}
\label{eq:p1}
\hat{p}_{u, i}=\frac{\exp \left(\frac{1}{|\mathcal{R}_{(i)}|}\sum_{r \in \mathcal{R}_{(i)}} s\left( \mathbf{e}^{\top}_r \hat{\mathbf{e}}_u, \hat{\mathbf{e}}_i\right)\right)}{\sum_{i^{\prime} \in \mathcal{N}_{(u)}} \exp \left(\frac{1}{|\mathcal{R}_{(i^{\prime})}|}\sum_{r \in \mathcal{R}_{(i^{\prime})}} s\left(\mathbf{e}^{\top}_r \hat{\mathbf{e}}_u, \hat{\mathbf{e}}_{i^{\prime}}\right)\right)}
\end{equation}
where $\hat{\mathbf{e}}_u$ and $\hat{\mathbf{e}}_i$ denotes knowledge-enhanced representation of $u$ and $i$, and ${e}_r$ denotes relation embedding. $\hat{p}_{u, i}$ embodies the personalized similarity of $u$ for each interacted $i$ according to the relations involved by $i$. Then the user preference can be acquired via the weighted sum of neighbors:
\begin{equation}
\hat{\mathbf{e}}_u=\hat{\mathbf{e}}_u+\sum_{i \in \mathcal{N}_{(u)}} m_{u, i} \hat{p}_{u, i} \hat{\mathbf{e}}_i
\end{equation}
where $m_{u, i}$ and $\hat{p}_{u, i}$ refer to a combination of hard and soft ways to remove or reduce the weight of unreliable edges. $\tilde{\mathbf{e}}_u$ can be obtained in the same way.
% Specifically, due to item representations from KG involving multi-relation semantics, and user preference mixing various information via message passing in the graph, we estimate the correlation degree between user $u$ and item $i$ across multiple relations. Each item has a relation set noted $\mathcal{R}_{(i)} = \{r | (h,r,t) \in \mathcal{T} \text{and } h \in \mathcal{I} \}$, and the similarity can be formulated as follows:

% \begin{equation}
% \label{eq:p1}
% \hat{p}_{u, i}=\frac{\exp \left(\frac{1}{|\mathcal{R}_{(i)}|}\sum_{r \in \mathcal{R}_{(i)}} s\left( \hat{e}_u {e}_r, \hat{e}_i\right)\right)}{\sum_{i^{\prime} \in \mathcal{N}_{(u)}} \exp \left(\frac{1}{|\mathcal{R}_{(i^{\prime})}|}\sum_{r \in \mathcal{R}_{(i^{\prime})}} s\left( \hat{e}_u {e}_r, \hat{e}_{i^{\prime}}\right)\right)}
% \end{equation}

% Where $\hat{p}_{u, i}$ denotes the semantic similarity between $u$ and $i$, $\hat{e}_u$ and $\hat{e}_{i^{\prime}}$ denotes semantic representation of $u$ and $i$, ${e}_r$ is relation embedding, and $\mathcal{N}_{(u)}$ is used to represent the set of neighbors of node $u$ in the user-item graph. $s(\cdot)$ denotes the inner product to estimate the similarity.

% The collaborative similarity is:
% \begin{equation}
% \label{eq:p2}
% \tilde{p}_{u, i}=\frac{\exp \left(s\left(\tilde{e}_u, \tilde{e}_i\right)\right)}{\sum_{i^{\prime} \in \mathcal{N}_(u)} \exp \left(s\left(\tilde{e}_u, \tilde{e}_{i^{\prime}}\right)\right)}
% \end{equation}

% Where $\tilde{e}_u$ and $\tilde{e}_i$ are additional user and item embedding, which is initialized to capture the pure collaborative signal.

\subsubsection{\textbf{Relation-aware Graph Self-enhancement.}} Due to the presence of noise, the aforementioned single-order denoising process is not sufficient to reduce the weight of the noisy edges and there is a chance of misclassification. Inspired by neighbor routing mechanism~\cite{icml19disgcn}, we design a relation-aware self-enhancement mechanism to generate augmented representation and correlation degree between users and items in mentioned two kinds of signals, which is formulated as:
\begin{equation}
\label{eq:iter1}
\hat{p}_{u, i}=\frac{\exp \left(\frac{1}{|\mathcal{R}_{(i)}|}\sum_{r \in \mathcal{R}_{(i)}} s\left(\mathbf{e}^{\top}_r \hat{\mathbf{e}}_u^{(n-1)}, \hat{\mathbf{e}}_i\right)\right)}{\sum_{i^{\prime} \in \mathcal{N}_{(u)}} \exp \left(\frac{1}{|\mathcal{R}_{(i^{\prime})}|}\sum_{r \in \mathcal{R}_{(i^{\prime})}} s\left(\mathbf{e}^{\top}_r \hat{\mathbf{e}}_u^{(n-1)}, \hat{\mathbf{e}}_{i^{\prime}}\right)\right)} 
\end{equation}
\begin{equation}
\label{eq:iter2}
\hat{\mathbf{e}}_u^{(n)}=\frac{\hat{\mathbf{e}}_u^{(n-1)}+\sum_{i \in \mathcal{N}_{(u)}} m_{u, i} \hat{p}_{u, i} \hat{\mathbf{e}}_i}{\left\|\hat{\mathbf{e}}_u^{(n-1)}+\sum_{i \in \mathcal{N}_{(u)}} m_{u, i} \hat{p}_{u, i} \mathbf{e}_i\right\|_2}
\end{equation}
where $m_{u,i}$ can be iteratively calculated by Eq.~(\ref{eq:binary_drop}). With conducting Eq.~(\ref{eq:iter1}) and~(\ref{eq:iter2}) $n$ times, noisy interactions will be gradually alienated and user representation $\hat{\mathbf{e}}_u^{(n)}$ is adjusted recursively towards the prototype of user preference~\cite{mm20grcn}. Collaborative user representation $\tilde{\mathbf{e}}_u$ can be generated in the same way.

If $\hat{p}_{u, i}$ and $\tilde{p}_{u, i}$ are both small, the corresponding interaction has little impact on the formation of user preference and can be considered as a soft-style denoising manner, while if the divergence between $\hat{p}_{u, i}$ and $\tilde{p}_{u, i}$ is significant and exceeds the threshold, hard-style denoising is triggered.

% 介绍 structure learning
% 介绍如何 aggregation

\subsection{Model Prediction}
We obtain the two pair representation of item $i$ and user $u$ at separate layers in different signals after $L$ layers, and then sum the multi-layer output as the final representation:
\begin{gather}\label{equ:final-representations}
    \hat{\mathbf{e}}_{v}=\sum\nolimits_{l=0}^L \hat{\mathbf{e}}_{v}^{(l)}, \qquad
    \tilde{\mathbf{e}}_{v}=\sum\nolimits_{l=0}^L \tilde{\mathbf{e}}_{v}^{(l)}
\end{gather}
where subscript $v$ denotes $u$ or $i$. By doing so, we segregate the complementary information of collaborative and knowledge semantics in the final representations, and we use cosine similarity to forecast how likely the user $u$ would engage with item $i$. Finally, the sum of the two-level similarity as the final prediction score $\hat{y}_{ui}$:
\begin{gather}\label{equ:prediction}
    \hat{y}_{u,i} = \cos(\tilde{\mathbf{e}}_u, \tilde{\mathbf{e}}_i) +  \cos(\hat{\mathbf{e}}_u, \hat{\mathbf{e}}_i)
\end{gather}

\subsection{Model Optimization}
\subsubsection{\textbf{Self-adapting loss function.}}
%  动态调整noisy ground-truth
Benefiting from explicitly pruning low-confidence interactions, we build a similarity bank $\mathcal{M}$ to dynamically collect and adjust the weight of each ``positive pair'' from implicit feedback during the training process. In addition, to alleviate the convergence problem, we opt for the contrastive loss~\cite{mao2021simplex} that introduces more negative samples and penalizes uninformative ones to optimize \ourmethod:
\begin{equation}
\mathcal{L}=\sum_{u, i \in \mathcal{D}}\left[m_{u, i}\left(1-\hat{y}_{u, i}\right)_{+} + \frac{1}{|\mathcal{N}|} \sum_{j \in \mathcal{N}}\left(\hat{y}_{u, j}-\beta\right)_{+}\right]
\end{equation}
where $m_{u,i} \in \mathcal{M}$ is the binary value derived from Section~\ref{sec:denoising_cf} to indicate whether each interaction $(u,i)$ should be retained during the training process, thus preventing the generation of harmful gradients that interfere with the user's real preference. The goal is to maximize the similarity of positive pairs while decreasing the similarity of negative pairs with a margin $\beta$. Moreover, $\mathcal{D}$ is the interaction data, $\mathcal{N}$ is the negative item set through random sampling from unobserved items with user $u$, and $(\cdot)_{+}$ is the ramp function $max(\cdot,0)$.

\subsubsection{\textbf{Indicators Gradient Estimation.}}
\label{sec:disarm}

An unbiased and low-variance gradient estimator DisARM~\cite{nips20disARM} is adopted to efficiently calculate the gradient of parameters $\alpha$. Let $M = (m_1,...,m_K)$ be a vector of $K$ independent Bernoulli variables with $m_i \sim \operatorname{Bern}(\sigma(\alpha_i))$, and the gradient of $\mathcal{\tilde{L}}$ in Eq.~\ref{eq:elbo} w.r.t $\alpha$ can be formulated as:
\begin{equation}
\nabla_{\alpha} \mathcal{\tilde{L}}(\alpha,\Theta)  =\frac{1}{2} \sum_i (f(\mathbf{b})-f(\mathbf{\tilde{b}}))((-1)^{\tilde{b_i}} \mathbbm{1}_{b_i \neq \tilde{b}_i} \sigma(|\alpha_i|))
\end{equation}
where $(\mathbf{b}, \mathbf{\tilde{b}}) = ((b_1,\tilde{b}_1),...,(b_K,\tilde{b}_K))^{T}$, discretized pair $(b_i, \tilde{b_i})=(\mathbbm{1}_{1-u_i<\sigma(\alpha_i)}, \mathbbm{1}_{u_i < \sigma(\alpha_i)})$, and $u \sim U(0,1)$ is sampled from Uniform distribution. $f(\mathbf{b})$ is the model loss obtained by setting each indicator $m_i$ to 1 if $1- u_i < \sigma(\alpha_i)$ in the forward pass of \ourmethod, 0 otherwise. The same strategy is applied to $f(\mathbf{\tilde{b}})$.

To this end, the gradient of binary indicators can be efficiently computed since: 1) Sampling from a Bernoulli distribution is replaced by sampling from a Uniform distribution between 0 and 1; 2) the estimator only involves two forward passes of model to calculate gradient, which can easily achieve with training.

\subsection{Model Analysis}
\subsubsection{\textbf{Model Size.}} The model parameters of \ourmethod consist of (1) trainable stochastic binary masks $\{\alpha_i | \alpha_i\in \Set{\alpha}\}$; (2) ID embeddings of users, items, relations, and other KG entities $\{\hat{\Mat{e}}_u,\hat{\Mat{e}}_i,\tilde{\Mat{e}}_u,\tilde{\Mat{e}}_i,\Mat{e}_e,\Mat{e}_r|u\in\Set{U},i\in\Set{I},e\in\Set{E},r\in\Set{R}\}$; and (3) two transformation parameters $W_{(1)}, W_{(2)}$ for compositional knowledge-refined aggregation.

\subsubsection{\textbf{Time Complexity.}} The time cost of \ourmethod mainly comes from two components: stochastic binary masks, aggregation, and self-enhancement schemes. The complexity of the stochastic binary masks are from DisARM, which requires two-forward pass of network and it's much less expensive than standard gradient backpropagation. In the aggregation over KG, $O(|\Set{G}_k|dL)$ is required to update entity representations, where $\Set{G}_k$, $d$, $L$ denote the number of KG triplets, the embeddings size, and the number of layers. In the user-item graph aggregation and self-enhancement schemes, the computational complexity of user and item embeddings in dual information signals is $O(2|\Set{G}_b|dLn)$, where $\Set{G}_b$ denotes the number of interactions, and $n$ is the iteration times. \ourmethod achieves comparable complexity to state-of-the-art knowledge-enhanced recommendation models.

\section{Experiments}
\label{sec:expperiments}
We present empirical results to demonstrate the effectiveness of our proposed \ourmethod framework. The experiments are designed to answer the following three research questions:
\begin{itemize}[leftmargin=*]
    \item \textbf{RQ1:} How does \ourmethod perform, compared with the state-of-the-art knowledge-aware recommendation models and denoising recommendation models?
    \item \textbf{RQ2:} How do different components of \ourmethod (\ie adaptive knowledge refining, contrastive denoising learning, and the depth of propagation layers) and the noise in interactions and KG affect the performance of \ourmethod?
    \item  \textbf{RQ3:} Can \ourmethod give intuitive impression of denoising results?
\end{itemize}

\subsection{Experimental Settings}
\label{sec:experimental_settings}

\begin{table}[t]
    \vspace{-3pt}
    \caption{Statistics of the datasets.}
    \vspace{-10pt}
    \label{tab:dataset}
    \resizebox{0.47\textwidth}{!}{
    \begin{tabular}{c|l|r|r|r}
    \hline
    \multicolumn{1}{l|}{} &  & \multicolumn{1}{c|}{Alibaba-iFashion} & \multicolumn{1}{c|}{Last-FM} & \multicolumn{1}{c}{Yelp2018}  \\ \hline\hline
    \multirow{3}{*}{\begin{tabular}[c]{@{}c@{}}User-Item\\ Interaction\end{tabular}} & \#Users & 114,737 & 23,566 & 45,919 \\
     & \#Items & 30,040 & 48,123 & 45,538  \\
     & \#Interactions & 1,781,093 & 3,034,796 & 1,185,068  \\ \hline\hline
    \multirow{2}{*}{\begin{tabular}[c]{@{}c@{}}Knowledge\\ Graph\end{tabular}} & \#Entities & 59,156 & 58,266 & 90,961  \\
     % & \#CL-Triplets & 105,470 & 369,179 & 869,611  \\
     % & \#SL-Triplets & 173,685 & 95,388 & 984,093  \\
     & \#Total-Triplets & 279,155 & 464,567 & 1,853,704  \\ \hline
    \end{tabular}}
    \vspace{-10pt}
\end{table}

\subsubsection{\textbf{Dataset Description.}}\label{sec:dataset}
We conduct experiments on three benchmark datasets: Alibaba-iFashion, Last-FM, and Yelp2018.
\begin{itemize}[leftmargin=*]
    \item \textbf{Alibaba-iFashion~\cite{kdd19pog}.} This is a fashion outfit dataset collected from Alibaba's online shopping system, which contains user-outfit click history, and outfits are viewed as items. Each outfit consists of several fashion staffs (\eg shoes, tops), and each staff has different fashion categories (\eg sweater, T-shirt).
    \item \textbf{Yelp2018\footnote{\url{https://www.yelp.com/dataset}}.} This is a local business rating dataset collected by Yelp. We use the 2018 edition dataset of the Yelp challenge, where local businesses like restaurants and bars are viewed as the items.
    \item \textbf{Last-FM\footnote{\url{https://grouplens.org/datasets/hetrec-2011}}.} This is a music listening dataset collected from Last.fm music website, where the tracks are viewed as items. We take the subset of the dataset where the timestamp is from Jan 2015 to June 2015.
\end{itemize}

Following previous work~\cite{kdd19kgat, www21kgin}, we collect the two-hop neighbor entities of items in KG to construct the item knowledge graph for each dataset. We use the same data partition with~\cite{kdd19kgat, www21kgin} for comparison (\ie the proportions of training, validation, and testing set are 80\%, 10\%, and 10\% for all datasets). Table~\ref{tab:dataset} presents the overall statistics of the three datasets used in our experiments. Meanwhile, in order to evaluate the denoising capability of \ourmethod, we follow~\cite{sigir2022rgcf} to construct three corresponding polluted datasets by adding noise into three real-world datasets (denoted as Polluted Alibaba-iFashion, Polluted Last-FM, Polluted Yelp2018). Specifically, for each original dataset, we randomly drop the observed user-item interaction that $\mathbf{R}_{ui} = 1$ and sample an item that the user has not adopted before as noisy interaction to replace the dropped one. The rate of the replaced observed interactions is set to 5\% by default, and we only inject noise in the training and validation sets.

\subsubsection{\textbf{Evaluation Metrics.}}
We adopt the all-ranking strategy to evaluate the performance~\cite{www21kgin}. In the test set, we regard all the items that user has not interacted with before as negative samples. To evaluate the performance of top-$N$ recommendation, we use adopt two widely-used evaluation metrics~\cite{kdd19kgat, www21kgin} Recall@$N$ and NDCG@$N$. We report the average results across all users in the test set with $N$ = 20 by default.

\subsubsection{\textbf{Baselines.}}
For performance evaluation, We compare \ourmethod with various baselines, including KG-free (MF), embedding-based (CKE), propagation-based methods(KGNN-LS, KGAT, CKAN, KGIN, MCCLK, KGCL). Besides, we also include robust recommendation models (T-CE, SGL, SGCN, SimGCL) with pre-trained item representations from KG to demonstrate the denoising ability of \ourmethod.
\begin{itemize}[leftmargin=*]
    \item \textbf{MF}~\cite{uai09BPRloss} is a benchmark factorization model, which only considers the user-item interactions and leaves KG untouched. 
    % Here, we use ID embeddings of users and items to perform the prediction.
    \item \textbf{T-CE}~\cite{wsdm2021T-CE} is a state-of-the-art sample re-weighting method, which assigns zero or lower weight for large-loss samples on BCE loss to reduce the impact of noisy interaction.
    \item \textbf{SGL}~\cite{sigir2021sgl} is a state-of-the-art self-supervised graph recommendation, which constructs multiple graph views and then conducts contrastive learning for robust learning. We adopt SGL with Edge Dropout (ED) strategy.
    \item \textbf{SGCN}~\cite{sigir21mask} is a state-of-the-art graph structure learning method, which applies a stochastic binary mask to prune noisy edges.
    \item \textbf{SimGCL}~\cite{sigir22simgcl} is a state-of-the-art contrastive learning method, which proposes a simple contrastive strategy by adding uniform noises on embedding space to generate different views.
    \item \textbf{CKE}~\cite{kdd16cke} is a embedding-based method. It utilizes TransR~\cite{aaai15transR} to regularize item representations from KG, and feeds learned embeddings into MF framework.
    \item \textbf{KGNN-LS}~\cite{kdd19kgnn-ls} is a GNN-based method. It transforms KG into a user-specific graph and considers label smoothness, so as to generate personalized item embeddings.
    \item \textbf{KGAT}~\cite{kdd19kgat} is a propagation-based method, which combines the user-item graph with KG and recursively propagates the embeddings with an attention mechanism.
    \item \textbf{CKAN}~\cite{sigir20ckan} is based on KGNN-LS, which perform different aggregation mechanism on the user-item bipartite graph and KG respectively, to encode user and item representations.
    \item \textbf{KGIN}~\cite{www21kgin} is a propagation-based method, which models user-item interaction behaviors with different intents, and captures long-range semantics with a relation-aware aggregation scheme.
    \item \textbf{MCCLK}~\cite{sigir22mcclk} is based on contrastive learning, which considers multi-level graph view, including structural, collaborative semantic views to mine additional supervised signal.
    \item \textbf{KGCL}~\cite{sigir22kgcl} is a newly propagation-based method, which proposes a KG augmentation schema to guide a contrastive learning paradigm for robust recommendation.
\end{itemize}

\subsubsection{\textbf{Parameter Settings.}}
We implement \ourmethod in Pytorch, and have released our implementation to facilitate reproducibility. For a fair comparison, we fix the ID embedding size to 64, set the batch size to 4096, use the Xavier initializer~\cite{xavier2010} to initialize the model parameters, and optimize all models with Adam~\cite{Kingma2014Adam} optimizer. A grad search is applied for hyperparameters. We tune the learning rate among $\{10^{-4},10^{-3},10^{-2}\}$, the GNN layers $L$ in $\{1,2,3\}$ and the pruning threshold $\gamma$ among $\{0.1,...,0.5\}$. Besides, we set the number of negative samples $|\mathcal{N}|$ per user and the margin $\beta$ of loss function to $\{200,400,400\}$ and $\{0.6,0.7,0.8\}$ for Alibaba-iFashion, Last-FM, and Yelp2018 datasets, respectively. Moreover, we carefully tune the other parameters for all baseline methods by following the original papers to achieve optimal performance.

\begin{table*}[t]
\centering
\vspace{-15pt}
\caption{Overall performance comparison. ``${\dagger}$'' indicates the
improvement of the \ourmethod over the baseline is significant at the level of 0.01. The highest scores are in Bold. R and N refer to Recall and NDCG, respectively.
}
\vspace{-10pt}
\label{tab:overall-performance}
\resizebox{1.01\textwidth}{!}{
\begin{tabular}{c|c|c|c|c|c|c|c|c|c|c|c|c|c|c|c}
\hline
\multicolumn{1}{c|}{\textbf{Database}} & \multicolumn{1}{c|}{\textbf{Method}} & \multicolumn{1}{c|}{\textbf{MF}} & \multicolumn{1}{c|}{\textbf{T-CE}} &
\multicolumn{1}{c|}{\textbf{SGCN}} & \multicolumn{1}{c|}{\textbf{SGL}} & \multicolumn{1}{c|}{\textbf{SimGCL}} & \multicolumn{1}{c|}{\textbf{CKE}} &
\multicolumn{1}{c|}{\textbf{KGNN-LS}} & \multicolumn{1}{c|}{\textbf{KGAT}} & \multicolumn{1}{c|}{\textbf{CKAN}} & \multicolumn{1}{c|}{\textbf{KGIN}} &
\multicolumn{1}{c|}{\textbf{MCCLK}} & \multicolumn{1}{c|}{\textbf{KGCL}} & \multicolumn{1}{c|}{\textbf{\ourmethod}} & \multicolumn{1}{c}{\textbf{\%Imp.}}\\ \hline
\multirow{2}{*}{Alibaba-iFashion}                & R@20   & 0.1095$^{\dagger}$	& 0.1093$^{\dagger}$	& 0.1145$^{\dagger}$	& 0.1232$^{\dagger}$	& \underline{0.1243}$^{\dagger}$	& 0.1103$^{\dagger}$	& 0.1039$^{\dagger}$	& 0.1030$^{\dagger}$	& 0.0970$^{\dagger}$	& 0.1147$^{\dagger}$	& 0.1089$^{\dagger}$	& 0.1127$^{\dagger}$	  & \textbf{0.1372}	& 10.38\% \\
& N@20     & 0.0670$^{\dagger}$	& 0.0631$^{\dagger}$	& 0.0722$^{\dagger}$	& 0.0771$^{\dagger}$	& \underline{0.0780}$^{\dagger}$	& 0.0676$^{\dagger}$	& 0.0557$^{\dagger}$	& 0.0627$^{\dagger}$	& 0.0509$^{\dagger}$	& 0.0716$^{\dagger}$	& 0.0678$^{\dagger}$	& 0.0713$^{\dagger}$	  & \textbf{0.0879}	& 12.69\% \\ \hline 

\multirow{2}{*}{Yelp2018}                & R@20   & 0.0627$^{\dagger}$	& 0.0705$^{\dagger}$	& 0.0768$^{\dagger}$	& 0.0788$^{\dagger}$	& \underline{0.0799}$^{\dagger}$	& 0.0653$^{\dagger}$	& 0.0671$^{\dagger}$	& 0.0705$^{\dagger}$	& 0.0646$^{\dagger}$	& 0.0698$^{\dagger}$	& 0.0696$^{\dagger}$	& 0.0748$^{\dagger}$	  & \textbf{0.0842}	& 5.38\% \\
& N@20     & 0.0413$^{\dagger}$	& 0.0542$^{\dagger}$	& 0.0547$^{\dagger}$	& 0.0518$^{\dagger}$	& \underline{0.0520}$^{\dagger}$	& 0.0423$^{\dagger}$	& 0.0422$^{\dagger}$	& 0.0463$^{\dagger}$	& 0.0441$^{\dagger}$	& 0.0451$^{\dagger}$	& 0.0449$^{\dagger}$	& 0.0491$^{\dagger}$	  & \textbf{0.0544}	& 4.62\% \\ \hline 

\multirow{2}{*}{Last-FM}                & R@20   & 0.0724$^{\dagger}$	& 0.0814$^{\dagger}$	& 0.0863$^{\dagger}$	& 0.0879$^{\dagger}$	& 0.0824$^{\dagger}$	& 0.0732$^{\dagger}$	& 0.0880$^{\dagger}$	& 0.0873$^{\dagger}$	& 0.0812$^{\dagger}$	& \underline{0.0978}$^{\dagger}$	& 0.0671$^{\dagger}$	& 0.0686$^{\dagger}$	  & \textbf{0.1023}	& 4.60\% \\
& N@20     & 0.0617$^{\dagger}$	& 0.0683$^{\dagger}$	& 0.0759$^{\dagger}$	& 0.0775$^{\dagger}$	& 0.0736$^{\dagger}$	& 0.0630$^{\dagger}$	& 0.0642$^{\dagger}$	& 0.0744$^{\dagger}$	& 0.0660$^{\dagger}$	& \underline{0.0848}$^{\dagger}$	& 0.0603$^{\dagger}$	& 0.0629$^{\dagger}$	  & \textbf{0.0946}	& 11.56\% \\ \hline \hline 

\multirow{2}{*}{\shortstack{Polluted \\ Alibaba-iFashion}}                & R@20   & 0.0982$^{\dagger}$	& 0.0990$^{\dagger}$	& 0.1035$^{\dagger}$	& 0.1146$^{\dagger}$	& \underline{0.1161}$^{\dagger}$	& 0.0911$^{\dagger}$	& 0.0921$^{\dagger}$	& 0.0902$^{\dagger}$	& 0.0874$^{\dagger}$	& 0.1037$^{\dagger}$	& 0.0981$^{\dagger}$	& 0.1065$^{\dagger}$	  & \textbf{0.1312}	& 13.01\% \\
& N@20     & 0.0607$^{\dagger}$	& 0.0584$^{\dagger}$	& 0.0639$^{\dagger}$	& 0.0714$^{\dagger}$	&  \underline{0.0722}$^{\dagger}$	& 0.0630$^{\dagger}$	& 0.0471$^{\dagger}$	& 0.0542$^{\dagger}$	& 0.0448$^{\dagger}$	& 0.0643$^{\dagger}$	& 0.0613$^{\dagger}$	& 0.0672$^{\dagger}$	  & \textbf{0.0839}	& 16.20\% \\ \hline 

\multirow{2}{*}{\shortstack{Polluted \\ Yelp2018}}                & R@20   & 0.0589$^{\dagger}$	& 0.0669$^{\dagger}$	& 0.0697$^{\dagger}$	& 0.0755$^{\dagger}$	& \underline{0.0759}$^{\dagger}$	& 0.0634$^{\dagger}$	& 0.0612$^{\dagger}$	& 0.0642$^{\dagger}$	& 0.0609$^{\dagger}$	& 0.0679$^{\dagger}$	& 0.0667$^{\dagger}$	& 0.0718$^{\dagger}$	  & \textbf{0.0816}	& 7.51\% \\
& N@20     & 0.0392$^{\dagger}$	& 0.0477$^{\dagger}$	& 0.0480$^{\dagger}$	& 0.0492$^{\dagger}$	& \underline{0.0495}$^{\dagger}$	& 0.0412$^{\dagger}$	& 0.0401$^{\dagger}$	& 0.0407$^{\dagger}$	& 0.0416$^{\dagger}$	& 0.0436$^{\dagger}$	& 0.0422$^{\dagger}$	& 0.0472$^{\dagger}$	  & \textbf{0.0528}	& 6.67\% \\ \hline 

\multirow{2}{*}{\shortstack{Polluted \\ Last-FM}}                & R@20   & 0.0711$^{\dagger}$	& 0.0807$^{\dagger}$	& 0.0858$^{\dagger}$	& 0.0879$^{\dagger}$	& 0.0948$^{\dagger}$	& 0.0849$^{\dagger}$	& 0.0863$^{\dagger}$	& 0.0845$^{\dagger}$	&  0.0805$^{\dagger}$	& \underline{0.0960}$^{\dagger}$	& 0.0668$^{\dagger}$	& 0.0731$^{\dagger}$	  & \textbf{0.1053}	& 9.69\% \\
& N@20     & 0.0610$^{\dagger}$	& 0.0675$^{\dagger}$	& 0.0741$^{\dagger}$	& 0.0791$^{\dagger}$	& 0.0844$^{\dagger}$	& 0.0735$^{\dagger}$	& 0.0630$^{\dagger}$	& 0.0743$^{\dagger}$	&  0.0658$^{\dagger}$	& \underline{0.0849}$^{\dagger}$	& 0.0592$^{\dagger}$	& 0.0695$^{\dagger}$	  & \textbf{0.0988}	& 16.37\% \\ \hline 

\end{tabular}}
\vspace{-10pt}
\end{table*}

\begin{table}[t]
    \caption{Impact of knowledge refining \& denoising.}
    \centering
    \vspace{-10pt}
    \label{tab:ablation_study}
    \resizebox{0.465\textwidth}{!}{
    \begin{tabular}{l|c c |c c| c c}
    \hline
    \multicolumn{1}{c|}{\multirow{2}*{}}&
    \multicolumn{2}{c|}{Alibaba-iFashion} &
    \multicolumn{2}{c|}{Yelp2018} &
    \multicolumn{2}{c}{Last-FM} \\
      &recall & ndcg & recall & ndcg & recall & ndcg\\
    \hline
    \hline
    w/o AKR & 0.1317 & 0.0833 & 0.0826 & 0.0538 & 0.1008 & 0.0933 \\
    w/o CDL & 0.1240 & 0.0794 & 0.0801 & 0.0521 & 0.0984 & 0.0905 \\
    w/o AKR\&CDL & 0.1225 & 0.0773  & 0.0789 & 0.0512 & 0.0974 & 0.0903\\
    \hline
    \end{tabular}}
    \vspace{-5pt}
\end{table}

\begin{table}[t]
    \caption{Impact of the number of layers $L$.}
    \centering
    \vspace{-10pt}
    \label{tab:impact-of-layer-number}
    \resizebox{0.465\textwidth}{!}{
    \begin{tabular}{l|c c |c c| c c}
    \hline
    \multicolumn{1}{c|}{\multirow{2}*{}}&
    \multicolumn{2}{c|}{Alibaba-iFashion} &
    \multicolumn{2}{c|}{Yelp2018} &
    \multicolumn{2}{c}{Last-FM} \\
      &recall & ndcg & recall & ndcg & recall & ndcg\\
    \hline
    \hline
    \ourmethod-1 & 0.1356 & 0.0866 & 0.0840 & 0.0544 & 0.1017 & 0.0939 \\
    \ourmethod-2 & 0.1365 & 0.0873 & 0.0842 & 0.0545 & 0.1023 & 0.0946 \\
    \ourmethod-3 & 0.1372 & 0.0879 & 0.0841 & 0.0545 & 0.1021 & 0.0941 \\
    \hline
    \end{tabular}}
    \vspace{-5pt}
\end{table}

\begin{table}[t]
    \caption{Impact of the iteration times $n$.}
    \centering
    \vspace{-10pt}
    \label{tab:impact-of-iteration-times}
    \resizebox{0.455\textwidth}{!}{
    \begin{tabular}{l|c c |c c| c c}
    \hline
    \multicolumn{1}{c|}{\multirow{2}*{}}&
    \multicolumn{2}{c|}{Alibaba-iFashion} &
    \multicolumn{2}{c|}{Yelp2018} &
    \multicolumn{2}{c}{Last-FM} \\
      &recall & ndcg & recall & ndcg & recall & ndcg\\
    \hline
    \hline
    n-1 & 0.1344 & 0.0852 & 0.0826 & 0.0527 & 0.1009 & 0.0929 \\
    n-2 & 0.1369 & 0.0875 & 0.0842 & 0.0545 & 0.1023 & 0.0946 \\
    n-3 & 0.1372 & 0.0879 & 0.0841 & 0.0543 & 0.1023 & 0.0945 \\
    \hline
    \end{tabular}}
    \vspace{-10pt}
\end{table}

\subsection{Performance Comparison (RQ1)}
\label{sec:performace}
We begin with the performance comparison \wrt Recall@20 and NDCG@20. The experimental results are reported in Table~\ref{tab:overall-performance}, where \%Imp. denotes the relative improvement of the best performing method (starred) over the strongest baselines (underlined). We have the following observations:
\begin{itemize}[leftmargin=*]
    \item \ourmethod consistently yields the best performance on all the datasets. In particular, it achieves significant improvement even the strongest baselines \wrt NDCG@20 by 12.69\%, 4.62\%, and 11.56\% in Alibaba-iFashion, Yelp2018, and Last-FM, respectively. We attribute these improvements to the fine-grained modeling and denoising collaborative learning of \ourmethod: (1) By pruning irrelevant facts from KG with parameterized masks and introducing multi-channel information aggregation scheme, \ourmethod is able to explicitly capture important triplets for recommendation and aggregating different knowledge associations. In contrast, all baselines ignore the different contributions of various facts in KG and simply use an aggregation scheme to propagate all kinds of knowledge associations. (2) Benefited from the contrastive denoising scheme, \ourmethod can explicitly prune noisy implicit interactions by contrasting semantic distance from both collaborative and knowledge aspects, while other denoising baselines (\eg T-CE, SGCN, and SGL) fail to leverage additional knowledge as denoising signals.
    \item Jointly analyzing the performance of \ourmethod across the three datasets, we find that the improvement on Alibaba-iFashion is more significant than that on other datasets. One possible reason is that the size of KG on Alibaba-iFashion is much smaller than that on Last-FM and Yelp2018, thus it is more important to mine useful knowledge information for modeling user preference.
    \item When the extra noise is injected into the training data, \ourmethod also yields significant improvement. Compared with the strongest baselines, \ourmethod achieves around 12\% performance improvement on three polluted datasets. For advanced knowledge-aware methods (\eg KGIN and KGCL), the injected noise significantly decreases their performances, while robust recommenders (\eg SGCN and SGL) are relatively less affected than their corresponding base models since they incorporate the denoising mechanism. Meanwhile, it is worth noting that most propagation-based methods are more sensitive to noise compared to MF and embedding-based methods. The reason is that message-passing scheme in GNN enlarges the negative impact of noise. \ourmethod removes irrelevant knowledge neighbors via applying learnable masks and prunes noisy interactions by contrasting collaborative and knowledge semantics, which can enhance the robustness and achieve better performances than other baselines. %SGL and SimGCL aims to mitigate the impact of noise with self-supervised learning, but it fails to tackle the fine-grained knowledge associations.
    \item Although the side information of KG is important to improve the explainability and accuracy of recommendations, we also find that existing robust recommendation models (\eg SGL and SimGCL) show competitive or even better performance compared with knowledge-aware recommendations (\eg KGIN and KGCL). One possible reason is that current knowledge-aware methods fail to fully explore the power of KG in a fine-grained manner, while our model is able to concentrate on clean triplets which are most useful for recommendation.
\end{itemize}

\begin{figure*}[t]
    \vspace{-15pt}
	\centering
	\includegraphics[width=1\textwidth]{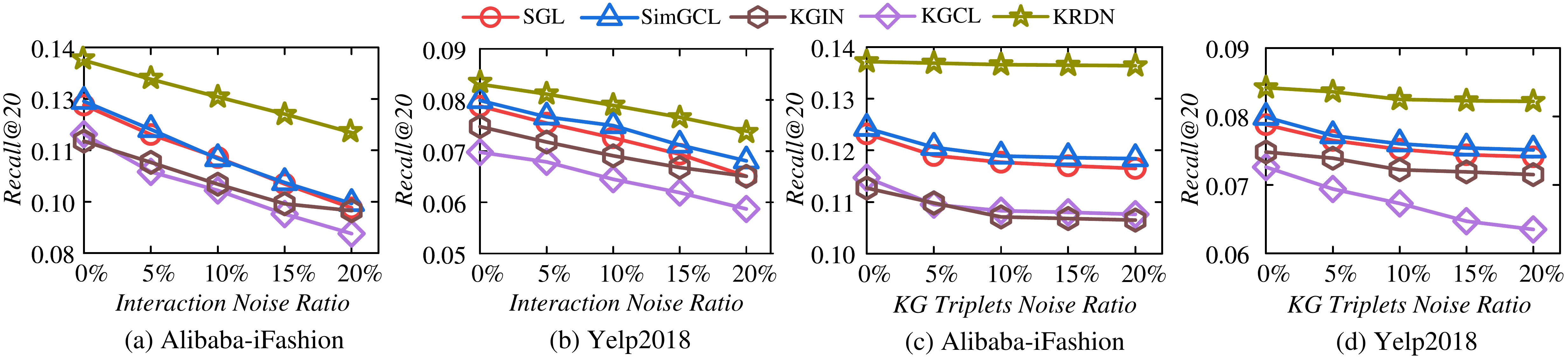}
        \vspace{-20pt}
	\caption{Impact of different ratio of noise in user-item graph and knowledge graph.}
	\label{fig:cf_kg_noise}
	\vspace{-10pt}
\end{figure*}

\subsection{Study of \ourmethod (RQ2)}
\label{sec:ablation_study}

As knowledge refining and denoising are at the core of \ourmethod, we conduct ablation studies to investigate the effectiveness. Specifically, how the presence of noisy interactions and facts, the adaptive knowledge refining, the contrastive denoising learning, the iteration times, and the number of propagation layers affect our model.

\subsubsection{\textbf{Robustness to Noisy Interactions and Facts.}}
We first conduct experiments to evaluate the robustness of \ourmethod with different ratios of noise. Following the dataset construction process described in Section~\ref{sec:dataset}, we pollute the training set and the validation set by replacing a certain ratio of original interactions with random noisy interactions, while keeping the testing set unchanged. Besides, we also add noise to the KG by randomly dropping tail entities $t$ and selecting new tail entities $t^\prime$ for these triplets.

Figure~\ref{fig:cf_kg_noise}(a) and figure~\ref{fig:cf_kg_noise}(b) show the experimental results (Recall@20) on polluted Aliababa-iFashion and Yelp2018 dataset. From the figure, we can observe that increasing the ratio of noisy interactions significantly reduces the performances of all baseline methods. The performance degradation of \ourmethod is much smaller than that of other methods, especially on Yelp2018 dataset. And the performance of \ourmethod is consistently better than baseline methods. The gap becomes larger when the ratio of noises increases from 0\% to 20\%. These observations further confirm the importance of denoising interactions in recommendation and demonstrate the robustness and effectiveness of \ourmethod.

In addition, figure~\ref{fig:cf_kg_noise}(c) and figure~\ref{fig:cf_kg_noise}(d) demonstrate the recommendation results on two datasets with noisy KG triplets. We find that \ourmethod is much more robust compared with other baselines, which remains nearly the same performance when the noisy triplets increase. However, other methods like KGCL show dramatic performance degradation, since they cannot differentiate the contribution of different facts in KG and also fail to prune irrelevant knowledge for recommendation tasks.

\subsubsection{\textbf{Impact of knowledge refining \& denoising.}}

We then verify the effectiveness of adaptive knowledge refining and contrastive denoising learning. To this end, three variants of \ourmethod are constructed by (1) removing the adaptive knowledge refining and contrastive collaborative denoising, called \ourmethod$_\text{w/o AKR\&CDL}$, (2) replacing the adaptive knowledge refining with simple single facet aggregation,  termed as \ourmethod$_\text{w/o AKR}$, and (3) discarding the contrastive denoising learning, named \ourmethod$_\text{w/o CDL}$. We summarize the results in Table~\ref{tab:ablation_study}.

Compared with the complete model of \ourmethod in Table~\ref{tab:overall-performance}, the absence of the adaptive knowledge refining and contrastive denoising learning dramatically degrades the performance, indicating the necessity of fine-grained KG modeling and denoising collaborative signals. Specifically, \ourmethod$_\text{w/o AKR\&CDL}$ directly aggregates all knowledge associations and ignores the noise in both KG and interactions, and thus, it fails to profile items properly and propagate information for learning use. Analogously, leaving the fine-grained knowledge associations unexplored (\ie \ourmethod$_\text{w/o AKR}$) also downgrades the performance. Although \ourmethod$_\text{w/o CDL}$ retains the fine-grained knowledge modeling for characterizing items, it is unable to provide discriminative signals for identifying user real behavior, incurring suboptimal user preference modeling.

\subsubsection{\textbf{Impact of Model Depth.}}
We also explore the impact of the number of aggregation layers. Stacking more layers is able to collect the high-order collaborative signals and knowledge associations for better capturing of the latent user behavior patterns but at a higher cost. Here, we search $L$ in the range of $\{1, 2, 3\}$, and report the results in Table~\ref{tab:impact-of-layer-number}. We have the following observations:
\begin{itemize}[leftmargin=*]
    \item Generally speaking, increasing the aggregation layers can enhance the performance, especially for Alibaba-iFashion datasets. We attribute such improvement to two reasons: (1) Gathering more relevant collaborative signals and knowledge association could provide informative semantics for learning high-quality representations, deepening the understanding of user interest. (2) The denoising module explicitly encodes both items’ profiles from KG and users' behaviors from interactions, which fully explores the power of KG for robust preference learning.
    \item It is worth mentioning that our model is less sensitive to the model depth, compared with other propagation-based methods~\cite{www21kgin,kdd19kgat,sigir22kgcl}. Specifically, \ourmethod could achieve competitive performance even when $L = 1$. This is because our fine-grained knowledge refining and denoising schemes can directly capture the most useful patterns from both user-item interactions and KG, while other methods need more layers to encode the latent semantics from the mixed and obscure information.
\end{itemize}

\subsubsection{\textbf{Impact of Self-enhancement Iteration Times.}}
To evaluate the effect of self-enhancement on contrastive denoising learning, we design experiments under different times of iteration. We search $n$ in the range of $\{1, 2, 3\}$, and report the result in Table~\ref{tab:impact-of-iteration-times}.

The performance of the model improves when the number of iterations is increased from 1 to 2. It indicates that iteratively executing the relation-aware graph self-enhancement is able to provide superior user profiles via adjusting the user representations towards the prototype of user preference and gradually reducing the weight of noisy neighbors. When we continue to increase the number of iterations, the model performance does not change much, which is probably due to the fact that we perform self-enhancement operations in each GNN layer, making the representations smooth.

\subsection{Case Study (RQ3)}
\label{sec:case_study}

\begin{figure}[t]
    \vspace{-10pt}
	\centering
	\includegraphics[width=0.40\textwidth]{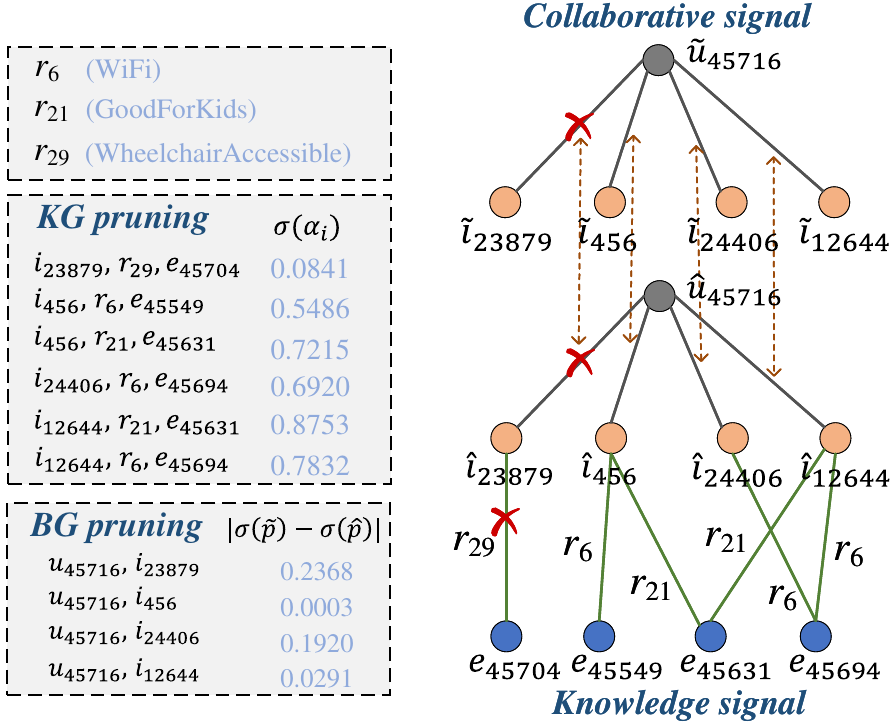}
	\vspace{-10pt}
	\caption{Explainability of denoising in Yelp2018. BG denotes a bipartite graph.}
	\label{fig:case_study}
	\vspace{-15pt}
\end{figure}

In this section, we present an example to provide an intuitive impression of \ourmethod's explainability. Toward that, we randomly select a user $u_{45716}$ from Yelp2018, and four directly connected items and their associated entities. As shown in Figure~\ref{fig:case_study}, we find that:
\begin{itemize}[leftmargin=*]
\item Benefiting from knowledge refining in the knowledge graph with stochastic binary masks, we can intuitively infer the reason for some triplets that are regarded as task-irrelevant facts and are pruned. For example, the collaborative knowledge graph in Figure~\ref{fig:case_study} shows that the user prefers these businesses with \textit{WiFi} $(r_{6})$ and \textit{GoodForKids} $(r_{21})$, and the triplet $(i_{23879}, r_{29}, e_{45704})$ which contains the attribute of \textit{WheelchairAccessible} $(r_{29})$ is irrelevant to the user profile. Meanwhile, the model assigns the triplet $(i_{23879}, r_{29}, e_{45704})$ a lower probability value of 0.0841, so that it has a small chance of being selected and avoids affecting the modeling of the user profile.
%\item Using collaborative signals alone only captures structural similarities between nodes, which is incomplete. As shown in Figure~\ref{fig:case_study},  The $i_{23879}$ attribute exhibits semantics that is far from the $u_{45716}$ preferences. 
\item The foundation for pruning noise interactions can be clearly perceived. $u_{45716}$ and $i_{23879}$ interact with each other and thus have collaborative similarity, but $i_{23879}$ contains attributes (e.g., ``WheelchairAccessible'') that are far from $u_{45716}$'s preferences. Especially, Figure~\ref{fig:case_study} shows that the similarity of $(u_{45716}, i_{23879})$ has a great disagreement between collaborative and knowledge signals, so this interaction could be pruned when the threshold is set to 0.2. 
\end{itemize} 

\section{Related Work}
\label{sec:related_work}

\subsection{Knowledge-aware Recommendation}
Existing Knowledge-aware recommendation methods can be roughly grouped into three categories: embedding-based~\cite{kdd16cke, www18dkn,alg18cfkg,www19ktup},   path-based~\cite{kdd18metapath,aaai19explainable,kdd20hinrec} and propagation-based methods~\cite{cikm18ripple, kdd19kgat, kdd19kgnn-ls, www19kgcn, sigir20ckan, www21kgin, Du2022MetaKG, sigir22hakg, sigir22mcclk, sigir22kgcl}. \textbf{Embedding-based methods} learn entities and relations embeddings in KG via knowledge graph embedding (KGE) methods (\eg TransR~\cite{aaai15transR}) to strengthen the semantic representation in recommendation. For example, CKE~\cite{kdd16cke} utilizes TransR to learn the knowledge representation of items, and incorporates learned embeddings into matrix factorization (MF)~\cite{uai09BPRloss}. Although these methods reveal simplicity and flexibility via KGE, they fail to capture the long-range dependence of user-item relations. \textbf{Path-based methods} explore the long-range connectivity among users and items by constructing different semantic paths via KG. Those paths are used to predict user profiles with recurrent neural networks~\cite{aaai19explainable} or attention mechanism~\cite{kdd18metapath}. For instance, KPRN~\cite{aaai19explainable} extracts the meta-paths via KG entities and relations to model high-order relations of user-item interactions by RNNs. However, defining proper meta-paths is time-consuming for complicated knowledge graphs and inevitably leads to poor generalization for different recommendation scenarios~\cite{kdd20hinrec,kdd20metahin}. \textbf{Propagation-based methods} are inspired by the information aggregation mechanism of graph neural network (GNNs)~\cite{iclr17gcn,nips17sageGCN,iclr18gat,icml19SGCN,sigir19ngcf,sigir20lighGCN}, which iteratively integrate multi-hop neighbors into node representation to discover high-order connectivity. For instance, KGAT~\cite{kdd19kgat} constructs a collaborative knowledge graph (CKG) using user-item interactions and KG, then performs an attentive aggregation mechanism on it. KGIN~\cite{www21kgin} integrates long-range semantics of relation paths by a new aggregation scheme and disentangles user preference behind user-item interactions by utilizing auxiliary knowledge for better interpretability. Most recently, MCCLK~\cite{sigir22mcclk} and KGCL~\cite{sigir22kgcl} combine a contrastive learning paradigm and build cross-view contrastive frameworks as additional self-discrimination supervision signals to enhance robustness. However, most of them fail to consider the negative impacts of task-irrelevant triplets in KG. Our work can prune task-irrelevant knowledge associations and noisy implicit feedback simultaneously.

\subsection{Denoising Recommender Systems}
Considerable attention has been paid to the robustness of recommendation systems. Especially, implicit feedback could be vulnerable to inevitable noise and then degrade the recommendation performance~\cite{wsdm2021T-CE, www2022DeCA, mm2021IR, sigir2022rgcf, sigir22SGDL}. Some work has gone into dealing with the noisy implicit feedback problem. Sample selection is a simple idea, which selects informative samples and then trains the model with them~\cite{kdd2012WBPR, mm2021IR}. For instance, WBPR~\cite{kdd2012WBPR} assigns different sampling probabilities according to item prevalence. IR~\cite{mm2021IR} discovers noisy samples based on the difference between predictions and labels. Moreover, sample re-weighting is a valid class of methods~\cite{wsdm2021T-CE, www2022DeCA}. For example, T-CE~\cite{wsdm2021T-CE} considers that noisy examples would have larger loss values, and hence assigns lower weights to high-loss samples. Besides, some recent studies use auxiliary information~\cite{ijcai2020dfn} or design model-specific structures~\cite{sigir21mask, sigir2021sgl, sigir21enhanced} to achieve denoising. For instance, DFN~\cite{ijcai2020dfn} uses additional explicit feedback (\eg like and dislike) to extract clean information from noisy feedback. SGCN~\cite{sigir21mask} explicitly prunes the irrelevant neighbors in the message-passing stage through sparsity and low-rank constraints. Most recently, ~\cite{sigir2021sgl, www2022NCL, sigir22simgcl, sigir21clea} utilize contrastive learning as auxiliary supervision signals. For example, SimGCL~\cite{sigir22simgcl} constructs augmented views by adding uniform noise. However, little effort has been done toward performing explicit denoising by utilizing knowledge graphs.
\section{Conclusion}
\label{sec:conclusions}
%In this paper, we proposed a new KG-based robust recommendation method \ourmethod, which solves noise issues in KG and user-item bipartite graph simultaneously. A trainable stochastic binary mask and a compositional aggregation mechanism are applied to refine task-irrelevant knowledge in KG, and then an unbiased and low-variance gradient estimation is adopted to update the mask with downstream supervision signals. Besides, \ourmethod prunes low-confidence interactions in user-item bipartite graph through performing iteratively self-enhancement and comparing the difference of collaborative and knowledge signals. Extensive experiments on three real-world datasets have demonstrated the superiority of \ourmethod. In the future, we plan to investigate the dynamics of noise, because users' interests evolve over time and so do the patterns of noise, thus combining user interaction sequences and temporal knowledge graphs to be able to locate noise behavior more precisely. 

In this paper, we proposed a new knowledge-aware robust recommendation method \ourmethod, which solves noise issues in KG and user-item bipartite graphs simultaneously. \ourmethod can eliminate irrelevant semantics in the knowledge graph and reduce the interference of user history noise interaction (\eg wrong click, wrong purchase), so as to achieve more satisfactory personalized recommendations with excellent interpretability in various real-world scenarios. Extensive experiments on three real-world datasets have demonstrated the superiority of \ourmethod. In the future, we plan to investigate the dynamics of noise, because users' interests evolve over time and so do the patterns of noise, thus combining user interaction sequences and temporal knowledge graphs to be able to locate noise behavior more precisely.


\begin{acks}
This work was supported by the NSFC under Grants No. (62025206, 61972338, and 62102351). Yuren Mao is the corresponding author of the work.
\end{acks}

\bibliographystyle{ACM-Reference-Format}
\balance

\bibliography{refer}
\balance

\end{document}